\documentclass{elsart}

\usepackage{amssymb}
\usepackage{epsfig}

\begin{document}

\begin{frontmatter}



\title{The CLAS Electromagnetic Calorimeter at Large Angles}

%\author[label1,label2]{}
%\address[label1]{}
%\address[label2]{}

\author[Ge]{M. Anghinolfi},
\author[Lnf]{H. Avakian\thanksref{TJNAF}},
\author[Ge]{M. Battaglieri},
\author[Lnf]{N. Bianchi},
\author[Ge]{P. Corvisiero},
\author[Ge]{R. De Vita},
\author[Ge]{E. Golovach\thanksref{Mos}},
\author[Lnf]{V. Gyuriyan\thanksref{TJNAF}},
\author[Lnf]{M. Mirazita},
\author[Ge]{V. Mokeev\thanksref{TJNAF}},
\author[Lnf]{V. Muccifora},
\author[Ge]{M. Osipenko},
\author[Lnf]{E. Polli},
\author[Ge]{G. Ricco},
\author[Ge]{M. Ripani},
\author[Lnf]{F. Ronchetti},
\author[Lnf]{P. Rossi}
\author[Ge]{V. Sapunenko\thanksref{TJNAF}\corauthref{cor1}},
\ead{vvsap@jlab.org}
\author[Ge]{M. Taiuti\corauthref{cor1}}
\ead{taiuti@ge.infn.it}

\address[Ge]{Dipartimento di Fisica dell'Universit\`a di Genova or Istituto 
Nazionale di Fisica Nucleare, sezione di Genova, I-16146 Genova, Italy}

%\author[Lnf]{E. De Sanctis},
%\author[Lnf]{A.R. Reolon},

\address[Lnf]{Istituto Nazionale di Fisica Nucleare, 
Laboratori Nazionali di Frascati, I-00044 Frascati, Italy}

%\address[Jlab]{Thomas Jefferson National Accelerator Facility,
%Newport News, VA 23606, USA}

\corauth[cor1]{Corresponding authors} 

\thanks[Mos] {Permanent address: Institute of Nuclear Physics, 
Moscow State University, 119899 Moscow, Russia}

\thanks[TJNAF]{Presently address: Thomas Jefferson National Accelerator Facility,
Newport News, VA 23606, USA}

\begin{abstract}
The study of the response of the two modules of the large-angle electromagnetic shower calorimeter (LAC) of the CLAS detector to charged and neutral particles is reported. The results agree very well with the Monte Carlo simulation. 
The procedures adopted for the energy and timing calibration are also discussed, proving that the module geometry allows for simple self-calibrating energy and timing algorithms.
% Text of abstract
\end{abstract}

\end{frontmatter}

\def\umom{$GeV/c$}
\def\umomm{$GeV^{2}/c^{2}$}

% main text
\section{Introduction}

The accelerator CEBAF at Jefferson Lab provides up to $6 ~GeV$ electrons 
to three experimental halls. Hall B houses the Large Acceptance 
Spectrometer CLAS, which is designed to study multi-particle 
final-state reactions induced by photons and electrons at luminosities up
to $10^{34}$ $cm^{-2}$ $s^{-1}$. The charged particle momentum is analyzed using a 1 T toroidal magnetic field with azimuthal symmetry generated by six superconducting coils arranged around 
the beamline. Each sector between two coils acts as a single spectrometer instrumented with three layers of drift chambers (DC) for track reconstruction, one layer of scintillator counters for time-of-flight measurement (TOF), a gas filled Cherenkov counter (CC) and an electromagnetic shower calorimeter (EC).
Drift chambers and time-of-flight define the sensitive region of each sector in the range of the polar angle $\theta$ from $10^{\circ}$ to $150^{\circ}$, the forward calorimeter modules cover from 
$10^{\circ}$ to $45^{\circ}$ in each sector and the two modules of the large-angle calorimeter (LAC) extend the $\theta$ coverage up to $75^{\circ}$ in two CLAS sectors.
In 1998 the two LAC modules as a part of the CLAS detector have been commissioned 
and since then they have been successfully operating to run the Hall B experimental 
program with electron and photon beams. 
The main features of the LAC module have been already described in previous papers where we 
reported the properties of the main components as the plastic scintillators \cite{Tai95b,Ros96}, 
the light guides \cite{Tai96} and the photomultipliers \cite{Rip98}.
The detector structure and the response to minimum ionizing particle has been described in Ref.~\cite{Ang00}.
In this paper we report on the response of the LAC to charged and neutral particles. The results
are based on data taken with both an electron and a photon beam impinging on a cryogenic target filled with liquid Hydrogen or liquid Deuterium.
We first discuss the energy and timing response obtained from the detection of pions and protons.
After that we show the performances for neutral particles: photons and neutrons.

\begin{figure}
\begin{center}
\epsfig{file=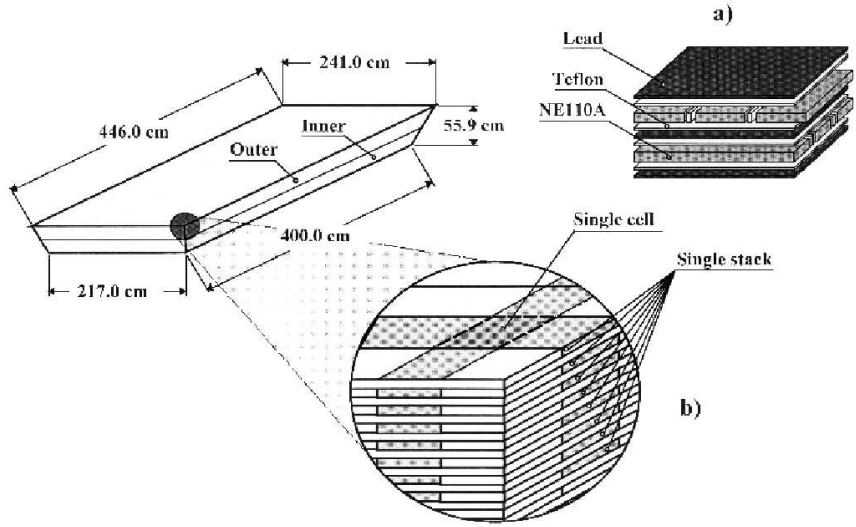,width=12 cm}
\caption{The conceptual drawing of the internal structure of the LAC module. 
\label{fig:Lacstructure}}
\end{center}
\end{figure}

\section{The LAC Energy Response}\label{section:EResponse}

To discuss the energy and timing responses, we first give a brief description of the detector. The conceptual drawing of the internal structure is shown in Fig.~\ref{fig:Lacstructure}: the LAC module has a rectangular shape with a sensitive area of 217x400 $cm^2$ and the internal structure is made of 33 layers, each composed of a 0.20 $cm$ thick lead foil and 1.5 $cm$ thick NE110A plastic scintillator bars. Each scintillator layer is protected from the contact with the lead by thin Teflon foils.
The width of the scintillators is roughly 10 $cm$ and is slightly increasing from the inner layers toward the outer layers to provide a focusing geometry. Scintillators in consecutive
layers are rotated by $90^{\circ}$ to form a 40x24 matrix of cells with area approximately 10x10 $cm^2$.
The module is vertically divided into an inner (17 layers) and an outer (16 layers) parts 
with separate light readouts. Scintillators lying one on top of the other with the same 
orientation form a stack. The focusing geometry is such that any straight trajectory
that originates from the target would hit two crossing stacks both in the inner part as well in the outer part.
For each stack the light is collected at both ends separately using light guides coupled to EMI 9954A photomultipliers. For each module there are 128 stacks and 256 photomultipliers.

\begin{figure}
\begin{center}
\epsfig{file=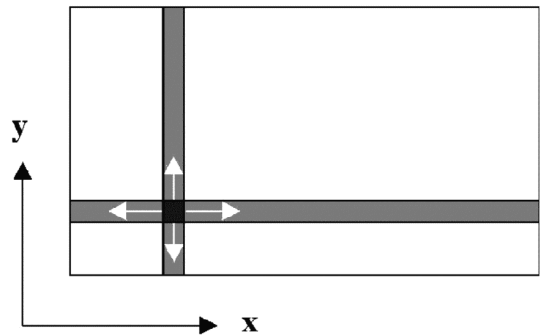,width=8 cm}
\caption{The cell selection procedure: minimum ionizing particles release energy in a region that is generally inside the dark grey square defined by the crossing of two orthogonal stacks (shown in light grey). The emitted light propagates along each stack and is collected at both end of the two stacks generating four ADC signals. 
\label{fig:CellSelection}}
\end{center}
\end{figure}

\subsection{The Energy Calibration}\label{section:ECalibration}

%The energy calibration is primarly based on the assumption that the response of photomultipliers
%and related electronics is linear. In addition, we used for each stack the value of the light attenuation length 
%previously measured \cite{Ang00}.
The energy calibration is performed continuosly checking off-line each photomultiplier.
The ADC pedestals are measured every day and the gain inhomogenities are corrected off-line setting the peak of the minimum ionizing particle (MIPs) distribution to the same value for all photomultipliers according to the following procedure already used during the tests with cosmic rays \cite{Ang00}. 
%value is stabilized changing the calibration constants after position correction.
%High momentum detected pions are used to equalize the calorimeter response. 
%All corrections are performed according to the procedure already used during the tests with cosmic 
%rays \cite{Ang00}. 
Minimum ionizing particles have enough kinetic energy to penetrate in the detector and exit from the back side.
The events corresponding to those particles (mainly high-momentum pions) that release their energy in a single cell are selected, in the off-line analysis, by imposing the presence of four ADC signals in the inner part and the four corresponding ADC signals in the outer part as shown in Fig.~\ref{fig:CellSelection}.
As a preliminary step, the correction for the dependence from the cell position is applied
considering that the light attenuation in scintillators relates the read-out charge $Q(x)$ to the ``true" value $Q_0$ according to the expression:
\begin{equation}\label{eqn:lightatten}
Q(x) = Q_0\left(e^{-x/\lambda}+\alpha e^{-(2L-x)/\lambda}\right),
\end{equation}
\noindent
where $x$ is given by the coordinates of the center of the cell, $L$ is the stack length and the two parameters $\lambda$ and $\alpha$ (representing, respectively, the light attenuation length and the reflectivity of the end surface of the scintillator bars) are the constants that have been determined for each stack by the cosmic ray calibration described in Ref.\cite{Ang00} and stored in a database.
After the position correction, the conversion factor charge-to-energy is evaluated, for each photomultiplier, in order to get the same MIPs peak position and to minimize the width of the overall deposited energy distribution.

After the energy calibration and the equalization of the read-out of the 512 photomultipliers, the calorimeter energy response has been studied for different detected particles and the results are discussed in the following sections.

\begin{figure}
\begin{center}
\epsfig{file=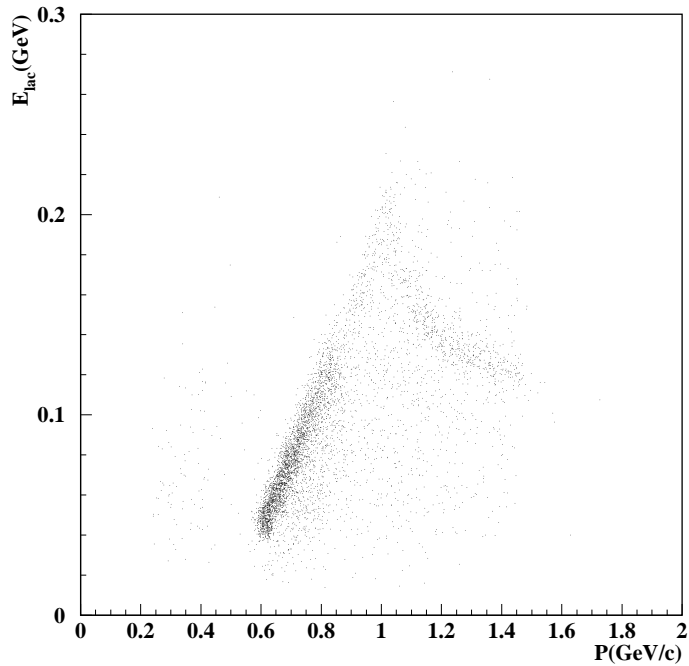,width=10 cm}
\caption{The energy deposited in the LAC module by protons .vs. the particle momentum $P$.
\label{fig:EvsP}}
\end{center}
\end{figure}

\begin{figure}
\begin{center}
\epsfig{file=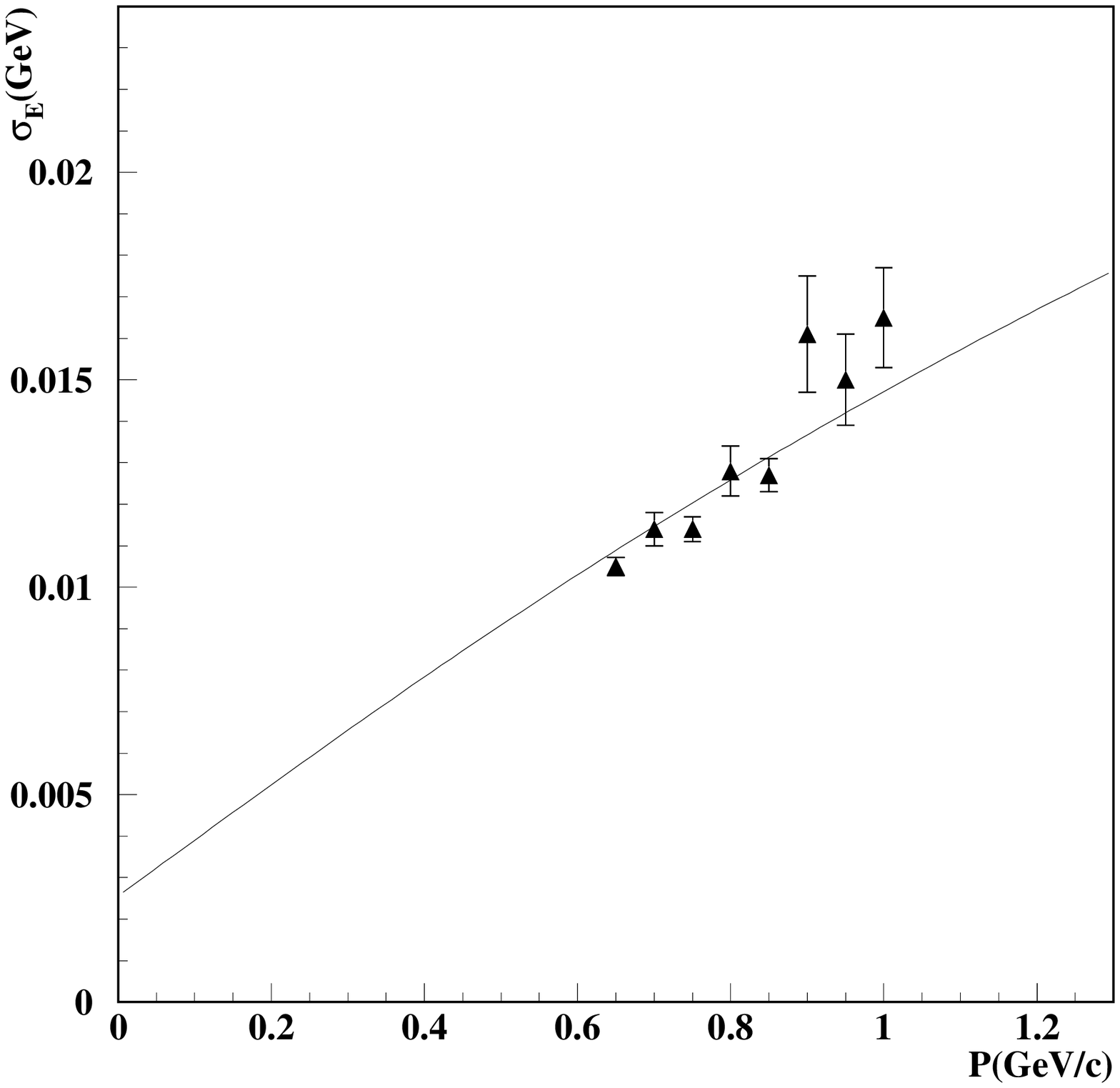,width=10 cm}
\caption{The deposited energy resolution vs. the particle momentum $P$ for the protons completely stopped in the calorimeter. The continuous line represents the result of the Monte Carlo simulation.
\label{fig:SigmavsP}}
\end{center}
\end{figure}

\subsection{The Response to the Ionizing Particles}\label{section:IParticles}

Protons are a powerful tool to check the linear response of individual stacks in a wide energy interval.
The protons detected by CLAS and used in this analysis originated from multipion emission off a liquid hydrogen target in electron scattering.
They have been selected by $\beta$.vs.momentum discrimination using informations from TOF and DC.
Fig.~\ref{fig:EvsP} shows a typical distribution of the deposited energy in the LAC as a function 
of the proton momentum.
Protons with momentum less than 0.6 \umom \ release all their kinetic energy in the TOF placed in front of the LAC scintillators as well as in the LAC front-housing and are not detected in LAC.
Protons with momentum up to 1 \umom \ are completely stopped by the calorimeter while higher-momentum protons release a fraction of their kinetic energy that approaches at very high momenta the asymptotic value of 106 $MeV$. This value corresponds to the energy released by MIPs in the LAC.

The momentum dependence of the deposited energy and the width of the deposited energy distribution agree with the Monte Carlo simulations. In Fig.~\ref{fig:SigmavsP} the momentum dependence of the width $\sigma_E$ of the deposited energy distribution of the protons stopped in the calorimeter is reported. 
The dependence of $\sigma_E$ from the deposited energy $E$ has been fitted with the expression
\begin{equation}\label{eqn:sigmaE}
\sigma_E = a + bE^{1/2} \;\; ,
\end{equation}
\noindent
with $E$ in $GeV$.
The parameter $a = (2.5 \pm 0.1)*10^{-3} \; GeV$ resulted very small while the value $b = (18.4 \pm 1.0)*10^{-3} \; GeV^{1/2}$ is larger than expected assuming the contribution only from the fluctuations in the light transmission and collection. Actually different effects contribute to the widening the deposited energy distribution; they include non uniformity in the effective scintillators and lead thicknesses due to different proton incident angle as well as non-uniformity in scintillator intercalibration. The contribution of each single effect cannot be easily disentangled, therefore we compared the overall response of the LAC modules with a realistic Monte Carlo simulation where we assumed 5 $photoelectron/MeV$ collected in average by each photomultiplier as already measured in cosmic ray tests \cite{Ang00}, and a photomultiplier intercalibration non-uniformity equal to 5 $\%$.
The continuous line in Fig.~\ref{fig:SigmavsP} represents the result of the simulation and the agreement is fairly good.

\subsection{The Response to the Electromagnetic Showers}\label{section:EShowers}

To study the response to electromagnetic showers, the only possibility is the detection of the photons from the $\pi^{0}$ decay, because the flux of electrons scattered at large angles is too low and the $\pi^{-}$ contamination too high.
The photons are identified selecting those events where the LAC energy signal has no TOF and DC signals in coincidence.
The size of the electromagnetic shower is normally larger than the stack width and several stacks 
are involved in the detection.
To reconstruct the photon energy first the two coordinates of the center of the shower are determined by weighting the coordinates of contiguous non-zero ADCs with their read-out, then 
for each ADC the read-out is corrected according to Eq.~\ref{eqn:lightatten}. Finally
the photon energy is obtained multiplying the sum of the corrected ADC read-outs
by the sampling factor $f = 0.32$. This factor has been determind from Monte Carlo simulations.
 
Unfortunately the photon energy spectra shows a continuous yield and cannot provide a direct test of the energy calibration. 
Nevertheless photon energy resolution could strongly affect the reconstructed mass of $\pi ^{0}$ and its investigation can provide an indirect check of this procedure as discussed in Sec.~\ref{section:pi0}. 

\section{The LAC Time Response}
The time response of the detector can affects the detection and identification of neutral particles. Photons originated from mesons decay can be disentangled from neutrons only using their different time-of-flight. To provide $\gamma -n$ discrimination good timing uniformity and resolution are required.
The detector geometry allows a self-calibrating procedure.
For each photomultiplier the relation between the particle time-of-flight $t$ and the corresponding TDC read-out $t'$ already corrected for the time-walk according to Ref.~\cite{Ang00} can be simply written as
\begin{equation}\label{eqn:timing}
t=l/\beta c=t'-\tau-x/v_{sc}-T_{0},
\end{equation}
being $l$ the length of the trajectory from the target to the detector, $\beta c$ the
particle velocity, $x$ the distance between the impact point in the detector 
and the photomultiplier, $v_{sc}$ the propagation velocity
of the light in the scintillator, $\tau$ the intrinsic time delay of each 
electronic chain and $T_{0}$ the overall calibration constant equal for all 
photomultipliers.
To provide the correct relationship between $t'$ and $\beta$, the 512 different $\tau$ 
constants and $T_{0}$ need to be set properly.

We first proceeded to the equalisation of the time response by setting the $\tau$
values using signal from MIPs. Only events hitting a single cell were selected imposing
the presence of four ADC signals in the inner part and of the four corresponding ADC 
signals in the outer part of the module. The ADC signals were originated from pairs of
photomultipliers coupled respectively to the right and to the left hand side of
the same scintillator stack. By using the corresponding TDC readouts $t'_{Left}$ and $t'_{Right}$
together with Eq.~\ref{eqn:timing} we evaluated for each stack the half-sum
\begin{eqnarray}\label{eqn:halfsum}
\Sigma t &=& \frac{1}{2} \left( t_{Right}+t_{Left} \right) \\ \nonumber
&=& \frac{t'_{Right}+t'_{Left}}{2}-\frac{\tau_{Right}+\tau_{Left}}{2}-L/2v_{sc}-T_{0},
\end{eqnarray}
being $L$ the scintillator lenght, and the half-difference
\begin{eqnarray}\label{eqn:halfdiff}
\Delta t &=& \frac{1}{2} \left( t_{Right}-t_{Left} \right) \\ \nonumber
&=& \frac{t'_{Right}-t'_{Left}}{2}-\frac{\tau_{Right}-\tau_{Left}}{2}.
\end{eqnarray}
With the adopted selection each event is represented by eight numbers: the values 
$\Sigma t_{S}$ and $\Delta t_{S}$ for the inner-short stacks, $\Sigma t_{L}$ and $\Delta t_{L}$ for
the inner-long stacks and similar values for the outer stacks.
$\Sigma t$ is simply the particle
time-of-flight and consequently, for each single event, the four $\Sigma t$ have to assume the same value, or equivalently, to satisfy the equation
\begin{equation} \label{eqn:deltasigma}
\Sigma t_{a} - \Sigma t_{b} = 0,
\end{equation}
for any pair $(a,b)$ of the $\Sigma t$ variables.

\begin{figure}
\begin{center}
\epsfig{file=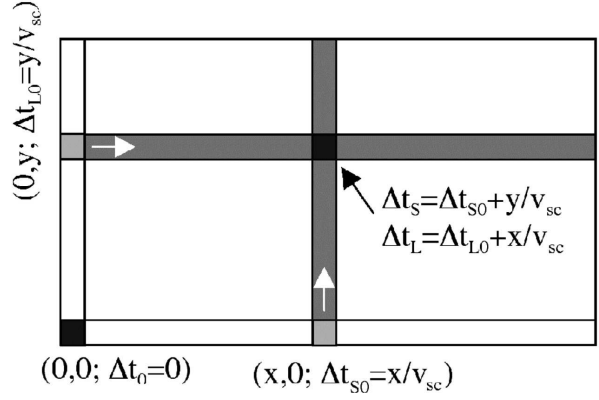,width=8 cm}
\caption{The $\Delta t$ timing calibration scheme. For a couple of crossing stacks the coordinates and the calibrated value of $\Delta t$ for the first cell are reported. The picture shows that with the assumed calibration, the $\Delta t$ readout for the cell located at the crossing point is the same for the two stacks.
\label{fig:DeltaTEqualization}}
\end{center}
\end{figure}

A similar relationship can be obtained for the four $\Delta t$ values as described in Fig.~\ref{fig:DeltaTEqualization}:
for each stack the $\Delta t$ value has been set such that for events in the first cell the correspondig timing would be
\begin{equation} \label{eqn:DeltaTCalibration}
\Delta t(0)=x/v_{sc},
\end{equation}
being $x$ the stack distance from the bottom-left angle of the module or equivalently the stack coordinate.
In this way it is easy to verify that the events detected in any cell with coordinates $(x,y)$ are associated to the same $\Delta t$ value for both short and long read-out:
\begin{equation} \label{eqn:DeltaTCalibration2}
\Delta t_S=\Delta t_L=\frac{x+y}{v_{sc}}.
\end{equation}
With this condition, for each event the four $\Delta t$ have to assume the same value, or equivalently
\begin{equation} \label{eqn:deltadelta}
\Delta t_{a} - \Delta t_{b} = 0,
\end{equation}
for any pair $(a,b)$ of the $\Delta t$ variables.

\begin{figure}
\begin{center}
\epsfig{file=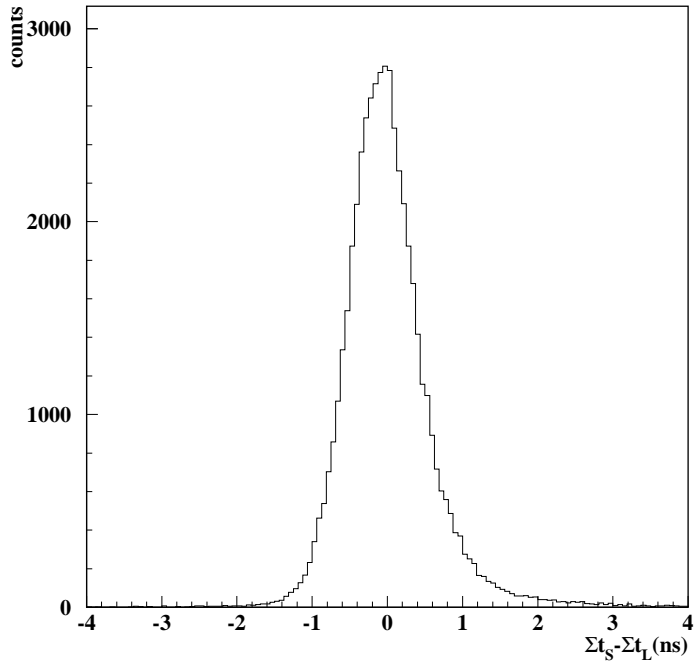,width=10 cm}
\caption{The distribution of the difference $\Sigma t_{S}-\Sigma t_{L}$ in the time-of-flight reconstructed with inner-short and inner-long stacks of the LAC for charged pions and protons detected in both modules.
\label{fig:deltaT}}
\end{center}
\end{figure}

Therefore it can be easily proved that equalizing the timing response of the LAC is equivalent to finding the values of the 512 $\tau$ that simultaneously set to zero the average value of the distributions 
$\left( \Sigma t_{S}-\Sigma t_{L} \right)$ and
$\left( \Delta t_{S}-\Delta t_{L} \right)$ separately for the inner and the outer part of the detector, minimizing at the same time their widths.
%The problem has been solved iteratively. 
%Considering as an example $\Sigma t$ first the mean value of the uncalibrated distribution is evaluated,
%then, alternatively for long and short stacks, the value of calibration constants $\frac{\tau_{Right}+\tau_{Left}}{2}$
%is set equal to the differ
In Fig.~\ref{fig:deltaT} a typical result is reported.

\begin{figure}
\begin{center}
\epsfig{file=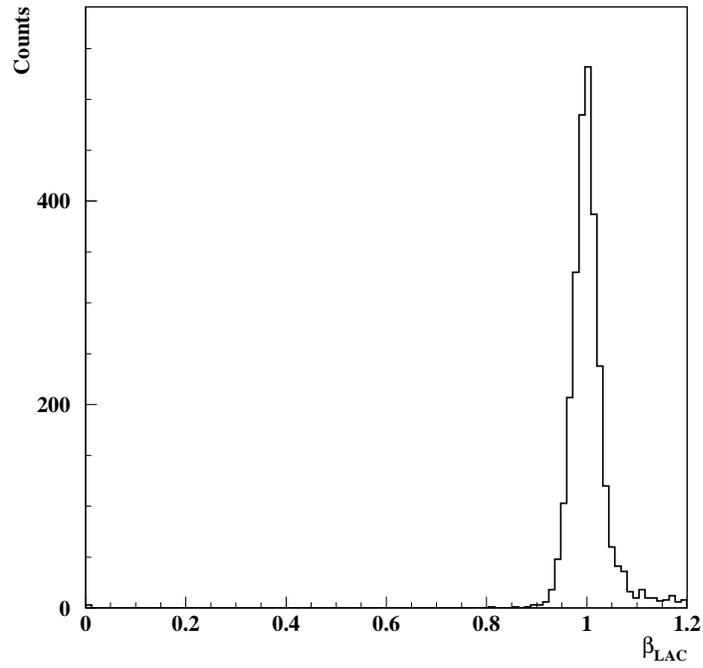,width=10 cm}
\caption{The reconstructed $\beta$ for photons detected in the LAC modules.
\label{fig:betagamma}}
\end{center}
\end{figure}

\begin{figure}
\begin{center}
\epsfig{file=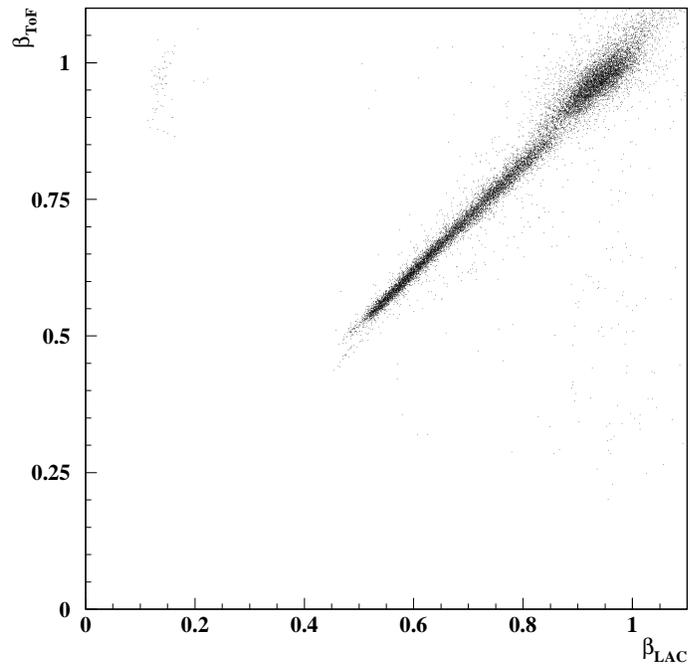,width=10 cm}
\caption{The time-of-flight reconstructed with TOF vs. LAC for charged pions and protons.
\label{fig:TOFvsLAC}}
\end{center}
\end{figure}

To fix $T_{0}$ we used the time-of-fligth of the photons originated from $\pi^0$ decay.
We selected photons candidates in the reaction $ep \rightarrow e'p X$ with missing mass 
equal to $M_{\pi ^{0}}$ and a neutral particle detected in coincidence by the LAC modules,
then we evaluated from the time-of-flight the corresponding $\beta = l/c\Sigma t$ values 
being $l$ the length of the straight trajectory from the target center to the reconstructed impact point in the LAC.
We set the $T_0$ value imposing the mean values of the $\beta$ distribution equal to 1.
In Fig.~\ref{fig:betagamma} the distribution of $\beta$ for detected photons shows a gaussian shape with $\sigma_{\beta} = 2.6\%$. 
Considering that the average distance from the target is 4 $m$ this value corresponds to a timing resolution $\sigma_t = 250 ~ps$ well in agreement with that obtained in the preliminary tests with cosmic rays \cite{Ang00}.
The timing response is very good in the whole $\beta$ range as shown in Fig.~\ref{fig:TOFvsLAC}
where the $\beta$ reconstructed with LAC is compared to the $\beta$ from the TOF 
for protons ($\beta < 0.85$) and $\pi^{+}$ ($\beta \approx 1$).
The data were taken using a 4.0 $GeV$ electron beam incident on a liquid H$_{2}$ target.

%\begin{figure}
%\begin{center}
%\epsfig{file=/work/pi0/vvsap/lac-publ/eps/theta-E.eps,width=10 cm}
%\epsfig{file=theta-E.eps,width=10 cm}
%\epsfig{file=/work/pi0/vvsap/lac-publ/eps/thetaEcut.eps,width=10 cm}
%\caption{Distribution of $\theta_{12}$ .vs. $E_{1}E_{2}$ for photons 
%from the reaction $ep \rightarrow e' \gamma \gamma X$ with at least one photon detected in LAC.
%\label{fig:ThetavsEE}}
%Continuous lines shows $\pi^{0}$ and $\eta$ position.
%\end{center}
%\end{figure}

%\begin{figure}
%\begin{center}
%\epsfig{file=/work/pi0/vvsap/lac-publ/eps/eta.eps,width=10 cm}
%\epsfig{file=eta.eps,width=10 cm}
%\caption{Reconstructed invariant mass for photons from the reaction $ep \rightarrow e' \gamma \gamma X$
% with at least one photon detected in LAC. 
%The continuous line is a fit with the sum of a gaussian that describes the $\eta$ peak 
%and an exponential background.
%\label{fig:etaMass}}
%\end{center}
%\end{figure}

\section{The Reconstruction of the $\pi^{0} \rightarrow 2\gamma$ Decay}\label{section:pi0}
The reconstructed value of the $\pi^{0}$ mass provides a quantitative check of the LAC energy equalisation and calibration. Any energy miscalibration would in fact provide an incorrect mean value 
and a larger width for the reconstructed mass distribution.
To verify that the energy calibration procedure discussed in Sec.~\ref{section:EResponse} is correct, we selected the events from the reaction $ep \rightarrow e' \gamma \gamma X$ with one photon detected in the LAC modules and one in the EC modules.
% To study the response of the LAC to the detection of the photons generated from
% $\pi ^{0}$ and $\eta$ decays, we used informations from other components of the CLAS detector.
% The photons were selected from the reaction $ep \rightarrow e' \gamma \gamma X$.
The neutral hits in the calorimeters were identified by the absence of a matched track in drift chambers. 
Photons detected by the LAC generally belongs to low momentum pions that decay emitting the two photons with a large relative angle $\theta_{12}$; therefore candidate events corresponding to the decay $\pi^{0} \rightarrow 2 \gamma$ were required to have a neutral hit in the LAC and a second neutral hit in the other LAC module or more favourably, in any EC module.

\begin{figure}
\begin{center}
\epsfig{file=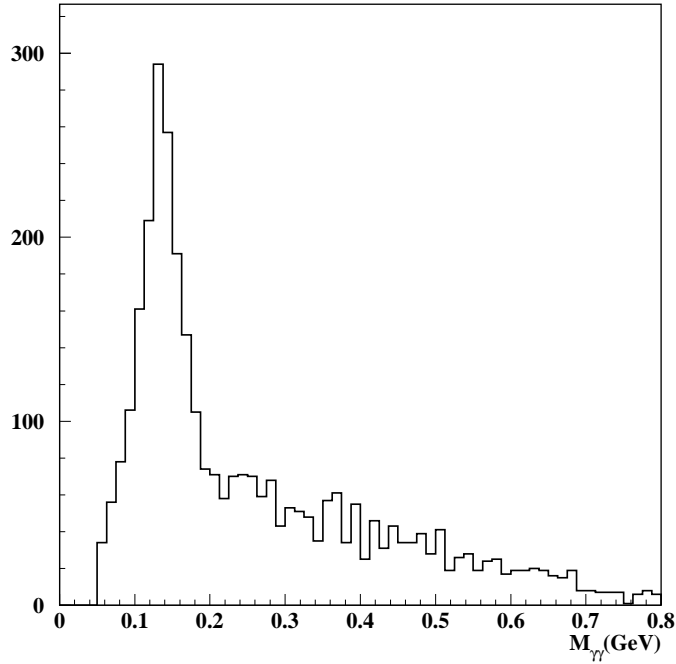,width=10 cm}
\caption{Reconstructed invariant mass for photons from the reaction 
$ep \rightarrow e' \gamma \gamma X$ with at least one photon detected in the LAC modules.
\label{fig:pi0Mass}}
\end{center}
\end{figure}

The meson mass has been reconstructed from the photon energies $E_{1}$, $E_{2}$
and the angle $\theta_{12}$ between the direction of the two photons according to
\begin{equation}\label{eqn:pi0mass}
M^2_{\gamma \gamma}=2 E_{1} E_{2} \left( 1-cos \theta_{12} \right).
\end{equation}
%Fig.~\ref{fig:ThetavsEE} shows for selected events the product  $E_{1}E_{2}$ versus the
%opening angle between the hits, evaluated using the reconstructed hit position.
%The data show a clear band of $\pi^{0} \rightarrow 2\gamma$ decays occurring within
%the invariant mass range 0.1-0.2 $GeV$ calculated using Eq.~\ref{eqn:pi0mass}.
%The data were taken using a 4.0 $GeV$ electron beam incident on a LH$_{2}$ target.
The reconstructed $M_{\pi^{0}}$, as shown in Fig.~\ref{fig:pi0Mass}, peaks at the right energy and the width of the distribution $\sigma = 28\%$ is well in agreement with the Monte Carlo simulations. 
The data were taken using a 4.0 $GeV$ electron beam incident on a liquid H$_{2}$ target.
 
%\begin{figure}
%\begin{center}
%\epsfig{file=/work/pi0/vvsap/lac-publ/eps/Theta-Mx.eps,width=10 cm}
%\epsfig{file=Theta-Mx.eps,width=10 cm}
%\caption{Azimuthal $\theta$ angle of the missing momentum direction versus
%missing mass for the reaction $ep \rightarrow e' \pi^+ X$.
%\label{fig:epi+ThetaVsMissMass}}
%\end{center}
%\end{figure}

%\begin{figure}
%\begin{center}
%\epsfig{file=/work/pi0/vvsap/lac-publ/eps/neutrons-lac.eps,width=10 cm}
%\epsfig{file=pics/neutrons-lac.eps,width=10 cm}
%\caption{Missing mass distribution for the reaction $ep \rightarrow e' \pi^+ X$.  The white histogram represents events with missing momentum directed toward the LAC modules, the grey histogram represents the same events with a hit in the LAC corresponding to a neutral particle with $\beta < 0.95$.
%\label{fig:gammapMissMass}}
%\end{center}
%\end{figure}

\section{The Neutron Detection}

The LAC detector has been designed to provide high neutron detection efficiency. 
From Monte Carlo simulations a detection efficiency larger than $20\%$ for neutron momenta higher than 0.7 $GeV/c$ is expected.
Neutrons as well as photons are identified by a neutral hit in the LAC and the particle velocity is reconstructed from the time-of-flight. Neutrons are separated from photons setting $\beta < 0.95$.

\begin{figure}
\begin{center}
\epsfig{file=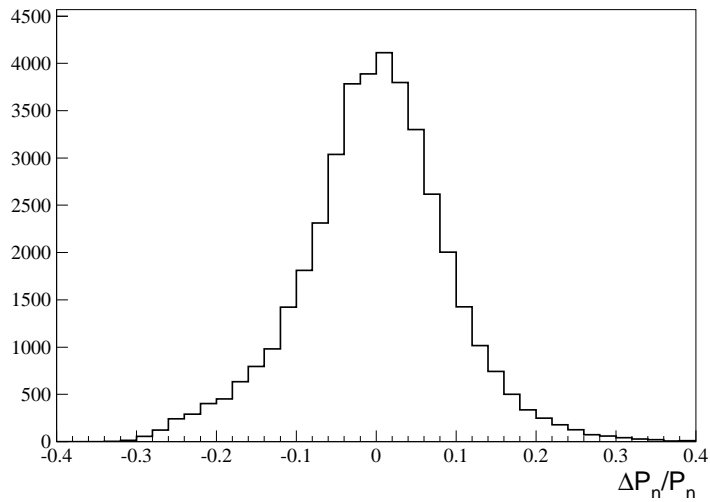,width=10 cm}
\caption{The relative difference between neutron momentum reconstructed with LAC time-of-flight
and the missing momentum for the reaction $\gamma d \rightarrow p X$.
\label{fig:nMomDis}}
\end{center}
\end{figure}

\begin{figure}
\begin{center}
\epsfig{file=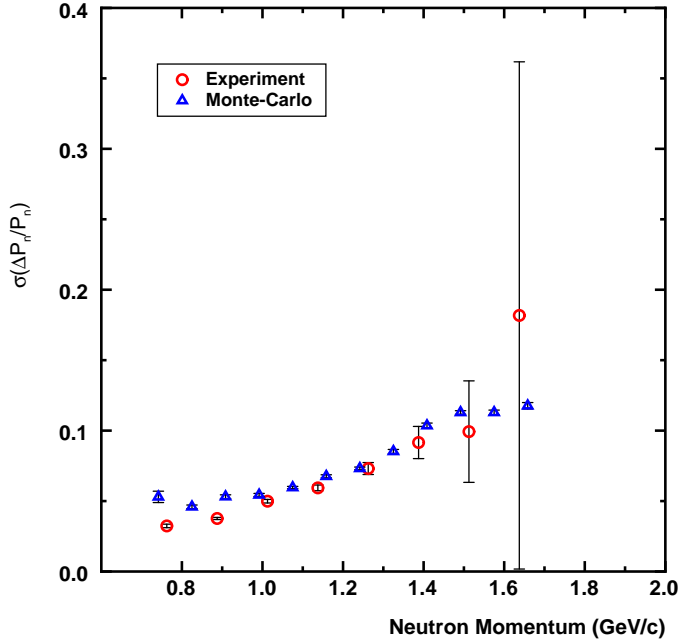,width=10 cm}
\caption{The relative width of the difference between the neutron momentum reconstructed with LAC time-of-flight and the missing momentum for the reaction $\gamma d \rightarrow p X$ as a function of $P_n$. Empty dots: experimental data; triangles: Monte Carlo simulations.
\label{fig:nMomRes}}
\end{center}
\end{figure}

To study the LAC response, the neutrons candidates were selected through the deuteron photo-disintegration $\gamma d \rightarrow p X$ imposing a cut in the missing mass distribution below the pion production threshold and the direction of the missing momentum oriented within the solid angle defined by the two LAC modules.
The data were taken using the 4.0 $GeV$ tagged photon beam incident on a liquid deuterium target. 
At this energy the neutrons detected in the LAC modules come mainly from the exclusive reaction $\gamma d \rightarrow p n$.
%satisfies these kinematical contraints as shown in fig.~\ref{fig:epi+ThetaVsMissMass}. 
%In fig.~\ref{fig:gammapMissMass} the missing mass $M_{X}$ for neutron candidates is reported.
%Neutrons detected by LAC are represented in fig.~\ref{fig:epi+MissMass} by the grey histogram. From the ratio of the two areas a detection efficiency of $50\%$ has been obtained.
Typical neutron momenta vary between 0.7 $GeV/c$ and 1.5 $GeV/c$.
The comparison between the momentum $P_n$ reconstructed by the LAC time-of-flight and the missing momentum $P_X$ for neutron candidates is reported in Fig.~\ref{fig:nMomDis}. The distribution of the relative difference $(P_n-P_X)/P_n$ is correctly centered on zero. The width of this distribution has been studied as a function of $P_n$ and compared to the Monte Carlo expectations. The results are reported in Fig.~\ref{fig:nMomRes} where the empty dots represent the experimental data and the full dots the simulation results: the agreement is fairly good in the whole investigated momentum range. 

\begin{figure}
\begin{center}
\epsfig{file=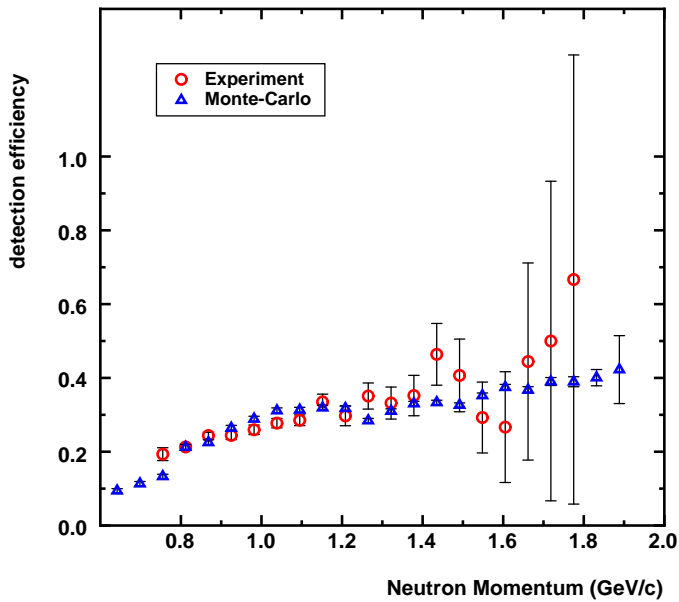,width=10 cm}
\caption{The neutron detection efficiency as a function of the neutron momentum. Empty dots: experimental data; Triangles: Monte Carlo simulations.
\label{fig:neff}}
\end{center}
\end{figure}

To evaluate the LAC efficiency we simply counted the fraction of candidates that generated a neutral hit in the LAC modules. In Fig.~\ref{fig:neff} we report the results (crosses) together with the Monte Carlo results (full dots) as a function of $P_n$: as for the momentum resolution, there is complete agreement in the whole investigated momentum range.

\section{Conclusions}

The two modules of the large angle calorimeter of the CLAS detector have been commissioned in 1998 and since then successfully operated. In this paper we have first discussed the calibration procedures proving that the detector geometry allows for simple energy and timing calibrating algorithms. These procedures are based on the detection of minimum ionizing particles (pions) that provide the relative energy and timing intercalibration. The absolute energy calibration for electromagnetic shower detection is obtained applying a correction factor as obtained from Monte Carlo simulations, and has been checked with the reconstructed $\pi ^0$ mass detecting one photon in
the LAC module and one photon in the forward calorimeter. 
The dependence of the resolution from the deposited energy has been studied with protons completely stopped in the calorimeter.
The timing absolute calibration is obtained setting $\beta = 1$ for the photons that originate from the $\pi ^0$ decay. The overall timing resolution for photons has been measured equal to $250~ps$. 
Finally the response to neutron has been studied as a function of the neutron momentum. The obtained momentum resolution and the detection efficiency fairly agrees with the Monte Carlo simulations. In particular the neutron detection efficiency is larger that $20\%$ for neutron momenta higher than 0.7 $GeV/c$.

Therefore we can conclude that the two LAC modules show the expected performances. 

\section{Aknowledgements}
This work was supported by the Istituto Nazionale di Fisica Nucleare. The Southeastern Universities Research Association (SURA) operates the Thomas Jefferson National Accelerator Facility for the United States Department of Energy under contract DE-AC05-84ER40150. We would like to thank E. Smith of the TJNAF for the fruitful discussions during the paper preparation.

% A useful Journal macro
\def\Journal#1#2#3#4{{#1} {\bf #2}, (#3) #4}

% Some useful journal names
\def\NCA{\em Nuovo Cimento}
\def\NIM{\em Nucl. Instrum. Methods}
\def\NIMA{{\em Nucl. Instr. and Meth.} A}
\def\NPB{{\em Nucl. Phys.} B}
\def\PLB{{\em Phys. Lett.}  B}
\def\PRL{\em Phys. Rev. Lett.}
\def\PRD{{\em Phys. Rev.} D}
\def\ZPC{{\em Z. Phys.} C}

\end{document}